# Investigation of Free Particle Propagator with Generalized Uncertainty Problem

*F. Ghobakhloo*[1] *and H. Hassanabadi* [1]

[1]Physics Department, Shahrood University of Technology, Shahrood, Iran P. O. Box: 3619995161-316, Shahrood, Iran

*Email:fatemeh.ghobakhloo@gmail.com

**Abstract**

We consider the Schrödinger equation with a generalized uncertainty principle for a free particle. We then transform the problem into a second ordinary differential equation and thereby obtain corresponding propagator. The result of ordinary quantum mechanics is recovered for vanishing minimal length parameter.

1. ## Introduction

The generalization of classical action principle into quantum theory appears in path integral formulation. Instead of a single classical path, the quantum version considers a sum, or better say, integral of infinite possible paths [1,2]. Although the main idea of path integral approach was released by N. Wiener, in an attempt to solve diffusion and Brownian problems, it was introduced in Lagrangian formulation of quantum mechanics of P. A. M. Dirac [1,2]. Nevertheless, the present comprehensive formulation is named after Feynman and extracted from his PhD thesis supervised by J. A. Wheeler [1, 2]. Feynman's formulation is now an essential ingredient in many fundamental theories of theoretical physics including quantum field theory, quantum gravity, high energy physics, etc [1,2,3].

On the other hand, we are now almost sure from fundamental theories such as string theory and quantum gravity that the ordinary quantum mechanics ought to be reformulated. In more precise words, a generalization of Heisenberg uncertainty principle, called Generalized Uncertainty Principle (GUP), should be considered at energies of order Planck scale [4-7]. This generalization corresponds to a generalization of wave equation of quantum mechanics. Till now, various wave equations of quantum mechanics, different interactions, as well as other related mathematical aspects and physical concepts have been considered in this framework [8-18].

In our note, we are going to combine these two subjects. Namely, we study the free particle propagator in Schrödinger framework in minimal length formulation. In section 2, we review the essential concepts of GUP and write the generalized Hamiltonian for free particle. In section 3, we obtain propagator for this system that the details of calculations are brought.

## 2. GUP-Corrected Hamiltonian

An immediate consequence of the ML is the GUP

$$\Delta x \geq \frac{\hbar}{\Delta p} + \alpha l_p^2 \frac{\Delta p}{\hbar} \qquad (1)$$

where the GUP parameter $\alpha$ is determined from a fundamental theory. At low energies, i.e. energies much smaller than the Planck mass, the second term in the right hand side of Eq. (1) vanishes and we recover the well-known Heisenberg uncertainty principle. The GUP of Eq. (1) corresponds to the generalized commutation relation

$$[x_{op}, p_{op}] = i\hbar\left(1 + \beta p^2\right) \qquad 0 \leq \beta \leq 1 \qquad (2)$$

where $x_{op} = x, p_{op} = p\left[1 + \beta(p)^2\right]$ and $0 \leq \beta \leq 1$. The limits $\beta \to 0$ and $\beta \to 1$ correspond to the standard quantum mechanics and extreme quantum gravity, respectively. Eq. (2) gives the minimal length in this case as $(\Delta x)_{\min} = 2l_p\sqrt{\alpha}$. It should be noted that in the deformed Schrödinger equation, the Hamiltonian does not have any explicit time dependence [19];

$$\left(\frac{p_{op}^2}{2m} + V(x)\right)\psi_n(x) = E_n\psi_n(x) \qquad (3).$$

This deformed momentum operator modifies the original Hamiltonian as

$$H = \frac{p^2}{2m} + V_{ff}(x) \qquad (4)$$

where

$$V_{ff}(x) = \beta\frac{p^4}{m} + V(x) \qquad (5)$$

The problem becomes much simpler if we consider [16];

$$p^2 = 2m\left(E_n^{(0)} - V(x)\right) \tag{6}$$

$$p^4 = 4m^2\left(E_n^{(0)} - V(x)\right)^2$$

In the free particle case, we therefore have

$$H = \frac{p^2}{2m} + 4\beta m \left(E_n^{(0)}\right)^2 \tag{7}$$

We now calculate the single free particle propagator corresponding to this deformed Hamiltonian in section 3.

## 3. Perspicuous Form of Propagator

If the wave function $\psi(x,t')$ is known at a time $t'$, we can explicitly write the wave function $\psi(x,t'')$ at a later time $t''$ using the propagation relation as [20]

$$\psi(x,t'') = \exp\left(-\frac{iH(t''-t')}{\hbar}\right)\psi(x,t') \tag{8}$$

For a small time interval $t'' - t' = \Delta t$, we have

$$\langle x | \exp\left(-\frac{iH\Delta t}{\hbar}\right) | x \rangle = \frac{1}{2\pi\hbar}\int dp\left(\langle x | \exp-\frac{i}{\hbar}\left(\frac{p^2}{2m}\Delta t\right) | p\rangle\langle p | \exp-\frac{i}{\hbar}\left(4\beta m \left(E_n^{(0)}\right)^2 \Delta t\right) | x\rangle\right) =$$

$$\int \frac{dp}{2\pi\hbar}\exp-\frac{i}{\hbar}\left(\frac{p^2}{2m} + 4\beta m \left(E_n^{(0)}\right)^2 \Delta t\right) \tag{9}$$

Therefore, the quantum mechanical propagator for small time interval $\Delta t = t - t'$, corresponding to this non-local Hamiltonian, can be written as

$$K\left(x'',t'';x',t'\right) = \int \exp\left(\frac{i}{\hbar}\int_{t'}^{t'+\Delta t} L(t)dt\right)\frac{dp}{2\pi\hbar} \tag{10}$$

In which the Lagrangian is given by [20]

$$L(t) = p \cdot \frac{(x''-x')}{(t''-t')} - \frac{p^2}{2m} - 4\beta m \left(E_n^{(0)}\right)^2 \tag{11}$$

Therefore, the propagator appears as

$$K(x'',t'';x',t') = \frac{1}{2\pi\hbar} \exp\frac{i}{\hbar}\left[\frac{(x''-x')^2 m}{2\Delta t} - 4\beta m\left(E_n^{(0)}\right)^2 \Delta t\right] \int dp.\exp\left[-\frac{i\Delta t}{2m\hbar}\left(p - \frac{(x''-x')m}{\Delta t}\right)^2\right] \quad (12)$$

or

$$K(x'',t'';x',t') = \frac{1}{2\pi\hbar} \exp\frac{i}{\hbar}\left[\frac{(x''-x')^2 m}{2\Delta t} - 4\beta m\left(E_n^{(0)}\right)^2 \Delta t\right] \int dU \exp\left(-\frac{i\Delta t}{2m\hbar}U^2\right) \quad (13)$$

with

$$U = \left(p - \frac{(x''-x')m}{\Delta t}\right)$$

In more explicit form, the propagator for free particle under minimal length is

$$K(x'',t'';x',t') = \left(\sqrt{\frac{m}{2i\pi\hbar\Delta t}}\right)\exp\frac{i}{\hbar}\left[\frac{(x''-x')^2 m}{2(t''-t')} - 4\beta m\left(E_n^{(0)}\right)^2 (t''-t')\right] \quad (14)$$

Now, if we assume $\beta = 0$, then the one dimensional free particle propagator is given by

$$K(x'',t'';x',t') = \left(\sqrt{\frac{m}{2i\pi\hbar\Delta t}}\right)\exp\frac{i}{\hbar}\left[\frac{(x''-x')^2 m}{2(t''-t')}\right] \quad (15)$$

In order to obtain free particle propagator for a finite time interval $(t''-t')$ we divide the interval into $N$ subintervals of equal length $\Delta t$ such that $(t''-t') = N\Delta t$. Now, the propagator of a finite time interval is written as

$$K(x'',t'';x',t') = \left(\sqrt{\frac{m}{2i\pi\hbar\Delta t}}\right)^N \int dx_1 dx_2 dx_3..dx_{N-1} \exp\frac{im}{2\hbar\Delta t}\left((x_1-x_0)^2 + (x_2-x_1)^2 + ...(x_N-x_{N-1})^2\right) \times$$

$$\exp\left(\frac{-i4\beta m}{\hbar}\left(E_n^{(0)}\right)^2 (t''-t')\right) \quad (16)$$

The intgral in Eq.(16) can be calculated as [20]

$$\int dx_1 dx_2 dx_3...dx_{N-1} \exp i\lambda\left[(x_1-x_0)^2+(x_2-x_1)^2+...(x_N-x_{N-1})^2\right]=$$

$$\frac{1}{\sqrt{N}}\left(\frac{i\pi}{\lambda}\right)^{\frac{N-1}{2}} \exp\left(\frac{i\lambda(x_N-x_0)^2}{N}\right) \qquad (17)$$

Substituting Eq.(17) into Eq. (16), the propagator is obtained as

$$K(x'',t'';x',t') = \left(\sqrt{\frac{m}{2i\pi\hbar\Delta t}}\right)^N \frac{1}{\sqrt{N}}\left(\frac{i\pi}{\lambda}\right)^{\frac{N-1}{2}} \exp\left(\frac{i\lambda(x_N-x_0)^2}{N}\right) \exp\left(\frac{-i4\beta m}{\hbar}\left(E_n^{(0)}\right)^2 \Delta t\right) \qquad (18)$$

where $\lambda = \frac{m}{2\hbar\Delta t}$. Replacing $x_N$ and $x_0$ by x″ and x′ respectively, and using $(t''-t')=N\Delta t$, we obtain the final expression as

$$K(x'',t'';x',t') = \left(\sqrt{\frac{m}{2i\pi\hbar\Delta t}}\right)^N \frac{1}{\sqrt{N}}\left(\frac{\lambda}{i\pi}\right)^{\frac{N-1}{2}} \exp\left(\frac{i\lambda(x_N-x_0)^2}{N}\right) \exp\left(\frac{-i4\beta m}{\hbar}\left(E_n^{(0)}\right)^2 \Delta t\right) \qquad (19)$$

Now, if we calculate the probability of detecting the particle at a finite region $\Delta x$, enclosing final point x″, from Eq. (19), we get

$$K(x'',t'';x',t') = \left(\sqrt{\frac{m}{2i\pi N\hbar\Delta t}}\right) \exp\left(\frac{im(x_N-x_0)^2}{2\hbar N \Delta t}\right) \exp\left(\frac{i4\beta m}{\hbar}\left(E_n^{(0)}\right)^2 \Delta t\right) \qquad (20)$$

In the limit $\beta \to 0$, the final form of propagator given by Eq. (21)

$$K(x'',t'';x',t') = \left(\sqrt{\frac{m}{2i\pi\hbar(t''-t')}}\right) \exp\left(\frac{im(x''-x')^2}{2\hbar(t''-t')}\right) \qquad (21)$$

which is the result in ordinary quantum mechanics.

**Conclusion**

We considered the nonrelativistic free particle propagation problem in an analytical manner in minimal length formalism. We first transformed arising differential equation into a second-order differential equation which included a modified effective potential. We next calculated the propagator. Apart from the application of the study, the work is of pedagogical interest in graduate physics.